\documentclass[conference]{IEEEtran}
\IEEEoverridecommandlockouts
\usepackage{cite}
\usepackage{amssymb,amsfonts,amsthm,amsmath}
\usepackage{algorithm}
\usepackage{algorithmic}
\usepackage{graphicx}
\usepackage{textcomp}
\usepackage{xcolor}
\usepackage{comment}
\usepackage{epstopdf}
\usepackage{subfigure} 
\newtheorem{remark}{Remark}
\newtheorem{theorem}{Theorem}
\def\BibTeX{{\rm B\kern-.05em{\sc i\kern-.025em b}\kern-.08em
    T\kern-.1667em\lower.7ex\hbox{E}\kern-.125emX}}
\begin{document}

\title{Channel Spreading Function-Inspired Channel Transfer Function Estimation for OFDM Systems with High-Mobility\\
}

\author{Yiyan~Ma,~\IEEEmembership{Member,~IEEE,}
        Bo~Ai*,~\IEEEmembership{Fellow,~IEEE,}
        Guoyu~Ma,~\IEEEmembership{Member,~IEEE,}
        Akram~Shafie,~\IEEEmembership{Member,~IEEE,}\\
        Qingqing~Cheng,~\IEEEmembership{Member,~IEEE,}
        Mi~Yang,~\IEEEmembership{Member,~IEEE,}
        Jingli~Li,~\IEEEmembership{Student Member,~IEEE,}\\
        Xuebo~Pang,~\IEEEmembership{Student Member,~IEEE,}
        Jinhong~Yuan,~\IEEEmembership{Fellow,~IEEE,}
        ~and~Zhangdui~Zhong,~\IEEEmembership{Fellow,~IEEE}
\vspace{-1em}
\thanks{Y. Ma, B. Ai, G. Ma, M. Yang, J. Li, X. Pang and Z. Zhong are with the Beijing Jiaotong University, Beijing, 100044, China. A. Shafie and J. Yuan are with the University of New South Wales, Sydney, NSW 2052, Australia. Q. Cheng is with Queensland University of Technology, Brisbane, QLD 4000, Australia. (Corresponding authors: boai@bjtu.edu.cn).}
}
\maketitle
\begin{abstract}
In this letter, we propose a novel channel transfer function (CTF) estimation approach for orthogonal frequency division multiplexing (OFDM) systems in high-mobility scenarios, that leverages the stationary properties of the delay-Doppler domain channel spreading function (CSF). First, we develop a CSF estimation model for OFDM systems that relies solely on discrete pilot symbols in the time-frequency (TF) domain, positioned at predefined resource elements. We then present theorems to elucidate the relationship between CSF compactness and pilot spacing in the TF domain for accurate CSF acquisition. Based on the estimated CSF, we finally estimate the CTF for data symbols. Numerical results show that, in high-mobility scenarios, the proposed approach outperforms traditional interpolation-based methods and closely matches the optimal estimator in terms of estimation accuracy. This work may pave the way for CSF estimation in commercial OFDM systems, benefiting high-mobility communications, integrated sensing and communications, and related applications.
\end{abstract}


\section{Introduction}
Orthogonal frequency division multiplexing (OFDM) has been widely implemented in 4G, WiFi, and 5G systems and is expected to remain essential in 6G and future mobile communication networks \cite{Andrews}. However, high-mobility scenarios with doubly (time and frequency) selective channels, anticipated in future 6G and beyond, present significant challenges for OFDM. In particular, Doppler spread can introduce inter-carrier interference (ICI) and cause fast channel fading, complicating the exact estimation of the channel transfer function (CTF), which represents the channel's time-frequency (TF) domain response \cite{b2}.\par

To support reliable CTF estimation, several estimators have been proposed for OFDM systems, including interpolation-based estimators \cite{b2}, minimum mean square error (MMSE) estimators \cite{b3}, and machine learning-based techniques \cite{b3}. However, the interpolation-based estimator may not be valid under fast-fading CTF, the MMSE estimator requires unavailable channel statistics, and machine learning approaches depend on extensive training data and may not generalize to different channel conditions \cite{b3}.\par

Recently, delay-Doppler (DD) multi-carrier (DDMC) schemes have been developed to enable reliable communication under high mobility \cite{Lin}, which are shown to provide greater reliability in high-mobility scenarios compared to OFDM \cite{weijiesurvey,raviteja,ziji2}. Notably, DDMC schemes leverage the channel spreading function (CSF), which characterizes the DD domain response of the channel. Unlike the CTF, which is typically considered invariant only within the coherence region, the CSF can be regarded as quasi-invariant in the stationary region larger than the coherence region \cite{matz}.\par

Thus, by leveraging the stationarity property of the CSF, it becomes feasible to estimate the CSF in OFDM systems using discrete pilots, followed by CTF estimation for data symbols based on the estimated CSF. While prior work has explored CSF estimation in OFDM systems \cite{b1,b111}, theoretical guidelines for configuring pilots in OFDM systems based on CSF compactness to achieve precise CSF estimation remain undeveloped.\par 

Motivated by these, this work presents a novel CSF-inspired CTF estimation approach for OFDM systems. We first develop a CSF estimation model based on predefined pilot symbols in the TF domain. Using Nyquist sampling principles, we then establish theorems that elucidate the relationship between pilot spacing and CSF compactness for reliable CSF estimation, and propose CSF estimation methods for both on-grid and off-grid Doppler cases. Based on the estimated CSF, we finally derive the CTF for data symbols in the TF domain. Simulations show that in high-mobility scenarios, the proposed approach surpasses interpolation-based methods and is comparable to the MMSE estimator in terms of accuracy, with lower complexity and without requiring channel statistical information.\par


{\bf Notation}: For the sake of brevity, denote ${\bf S}^\beta_{\alpha} = \{0, \beta, 2\beta, \cdots, \alpha-1\}$ as the positive integer set from $0$ to $\alpha-1$ with increment of $\beta$. For example, ${\bf S}^2_{N} = \{0, 2, 4,..., N-1\}$.

\section{System Model for CTF Estimation}
\subsection{Doubly Selective Channel Models for OFDM} 
The channel impulse response
(CIR) for a doubly selective channel can be modeled as follows \cite{matz}:
\begin{equation} 
h_{\rm TD}(t,\tau) = \sum^{P}_{i=1} h_i e^{j2\pi \nu_i t} \delta(\tau-\tau_i), 
\label{CIR_EQU1} 
\end{equation}
where $\tau \!\in\! [0, \!\tau_{\max}]$, $\nu \!\in\! [-\nu_{\rm max}, \nu_{\rm max}]$, and $h_i$ denote the delay, Doppler, and fading coefficient of the $i$-th multipath, respectively, $P$ represents the number of multipath, and $\tau_{\rm max}$ and $\nu_{\rm max}$ are maximum delay and Doppler, respectively.\par

We consider an OFDM system operating over a time duration of $NT$ and a bandwidth of $M\Delta f$, where $N$ and $M$ represent the number of OFDM symbols (or time slots) and subcarriers in each OFDM symbol, respectively. Let ${\bf x}_{\rm TF} \in {\mathbb C}^{M \times N}$ represent the transmitted symbols in the TF domain, and ${\bf x}_{\rm TF}[:,n]$ denote the $n$-th OFDM symbol with $n \in {\bf S}^1_N$. Using (\ref{CIR_EQU1}), the TF domain input-output relationship of OFDM systems under the doubly selective channel is \cite{Gaudio}:
{\begin{equation}
{\bf y}_{\rm TF}[:,n] = {\sum^{P}_{i = 1} h_i {\bf F} {\bf \Pi}^{l_i} {\bf \Lambda}_{n}^{(k_i)} {\bf F}^{H}}{\bf x}_{\rm TF}[:,n] + {\bf w},
\label{OFDMIO}
\end{equation}
where ${\bf y}_{\rm TF} \in {\mathbb C}^{M \times N}$ and ${\bf w} \in {\mathbb C}^{M \times N}$ denote the TF domain received symbols and additive white Gaussian noise (AWGN), respectively. Besides, ${\bf H}_{{\rm TF},n}=\sum^{P}_{i = 1} h_i {\bf F} {\bf \Pi}^{l_i} {\bf \Lambda}_{n}^{(k_i)} {\bf F}^{H} \in {\mathbb C}^{M \times M}$ represents the TF domain equivalent channel for the $n$-th OFDM symbol, ${\bf F} \in {\mathbb C}^{M \times M}$ is the discrete Fourier transform (DFT) matrix, and $k_i = NT \nu_i$ and $l_i = M\Delta f \tau_i$ are the normalized Doppler and delay, respectively. In this letter, we assume that $\tau_i$ for each path is on-grid, meaning $l_i$ is an integer \cite{raviteja}. Thus, the cyclic shift matrix corresponding to the delay $\tau_i$ becomes ${\bf \Pi}^{l_i}= \left( {\begin{array}{*{20}{c}}
{\bf 0}_{1\times (M-1)}&1\\
{\bf I}_{(M-1)\times (M-1)}&{\bf 0}_{(M-1)\times1}
\end{array}} \right)^{l_i}$. Additionally, ${\bf \Lambda}_{n}^{(k_i)} \in {\mathbb C}^{M \times M}$ is the phase shift matrix associated with the Doppler $\nu_{i}$ for the $n$-th OFDM symbol, i.e., ${\bf{\Lambda }}_n^{\left( {{k_i}} \right)} = {e^{j2\pi {\nu _i}nT}}{\rm{diag}}\left( {{e^{\frac{{j2\pi {k_i}(0)}}{{MN}}}},...,{e^{\frac{{j2\pi {k_i}(M - 1)}}{{MN}}}}} \right)$ \cite{raviteja}. 
We note that ${{\bf H}_{{\rm TF},n}}$ is not strictly diagonal due to ${\bf{\Lambda }}_n^{(k_i)}$, meaning that the Doppler of multipath introduces ICI. However, when $k_i \ll N$, for the purpose of simpler equalization, ${\bf{\Lambda }}_n^{(k_i)}$ can be approximated by ${e^{j2\pi {\nu_i}nT}}{{\bf I}_{M\times M}}$\footnote{This assumption is widely adopted, particularly in OFDM radar systems under high mobility conditions \cite{Gaudio}. Due to space limitations, we will explore the proposed approach considering ICI in detail in future research.}. Then the ICI can be neglected, allowing us to simplify (\ref{OFDMIO}) to:
\begin{equation}
{\bf y}_{\rm TF}[m,n] \approx {{h}_{{\rm TF}}[m,n]} {\bf x}_{\rm TF}[m,n] + {\bf w}[m,n],
\label{OFDMIOSIMPLE}
\end{equation}
where $m \in {\bf S}^{1}_{M}$ is the OFDM subcarrier index and $h_{\rm TF}[m,n]$ denotes the sampled version of the CTF $h(t,f)=\int h(t,\tau)e^{-j2\pi\tau f} d\tau$ \cite{matz}, and can be represented as:
\begin{equation}
h_{\rm TF}[m,n] = \sum^{P}_{i=1} h_i e^{j2\pi \nu_i nT} e^{-j2\pi \tau_i m\Delta f}.
\label{CTF}
\end{equation}

Besides, the CSF $h(\tau,\nu)=\int h(t,\tau)e^{-j2\pi \nu t}dt$ characterizes the DD domain channel \cite{matz}. The sampled version of $h(\tau,\nu)$, given by $h_{\rm DD}[k,l]$, can be obtained by applying the DFT along the time domain and the inverse DFT along the frequency domain on the CTF $h_{\rm TF}[m,n]$ in (\ref{CTF}), leading to:
\begin{equation}
\begin{aligned}
&\!\!h_{\rm DD}\left[{k},{l}\right]\! = DFT_{N}\{IDFT_{M}\{{{h}_{{\rm TF}}[m,n]}\}\}\\
&=\!\sum\limits_{i {=} 1}^P {{h_i}\sum\limits_{m {=} 0}^{M {-} 1} \!\frac{1}{\sqrt M}{{e^{ - j2\pi m\frac{{\left( {{l_i} - l} \right)}}{M}}}\!\sum\limits_{n {=} 0}^{N {-} 1} \!\frac{1}{\sqrt N}{{e^{j2\pi n\frac{{\left( {{k_i} - k} \right)}}{N}}}} } }.
\end{aligned}
\label{CSF}
\end{equation}
Here, $\forall k \in {\bf S}^{1}_{N}$ and $\forall l \in {\bf S}^{1}_{M}$ are indices for Doppler and delays, corresponding to resolutions $\frac{1}{NT}$ and $\frac{1}{M\Delta f}$, respectively.\par
\subsection{Proposed Pilot Arrangement Model for OFDM}
We note that $h_{\rm TF}[m,n]$ in (\ref{CTF}) for $NM$ resource elements (REs) cannot be directly obtained using the pilot structure typically adopted in OFDM systems, as only few pilots are placed at specific REs in TF domain \cite{b2}. To facilitate CTF estimation that is to be proposed in Section III, we define the intervals between adjacent pilots in the time and frequency domains as integers $d_t$ and $d_f$, respectively, with ${\rm mod}(N, d_t) = 0$ and ${\rm mod}(M, d_f) = 0$, as shown by Fig. \ref{FIG1} \cite{b2}. The choice of $d_t$ and $d_f$ will be discussed in Section III. We subsequently define the CTF corresponding to the OFDM frame with the proposed pilot arrangement as the {\it discrete CTF}, given by:
\begin{equation}
\begin{aligned}
&{h}^{\rm Discrete}_{\rm TF}[m,n] \!\!=\! \begin{cases} \!{h}^{\rm Pilot}_{\rm TF}[m,n],& \begin{aligned}& m\!\in\!{\bf S}^{d_f}_{M},n\!\in\!{\bf S}^{d_t}_{N}\end{aligned} \\
0, & {\rm elsewhere},
\end{cases}
\end{aligned}
\label{CTF_Dis}
\end{equation}
where ${h}^{\rm Pilot}_{\rm TF}[m,n]$ can be represented as in (\ref{CTF}).
\vspace{-0.45em}
\begin{figure}[t]
  \centering
  \includegraphics[width=.25\textwidth]{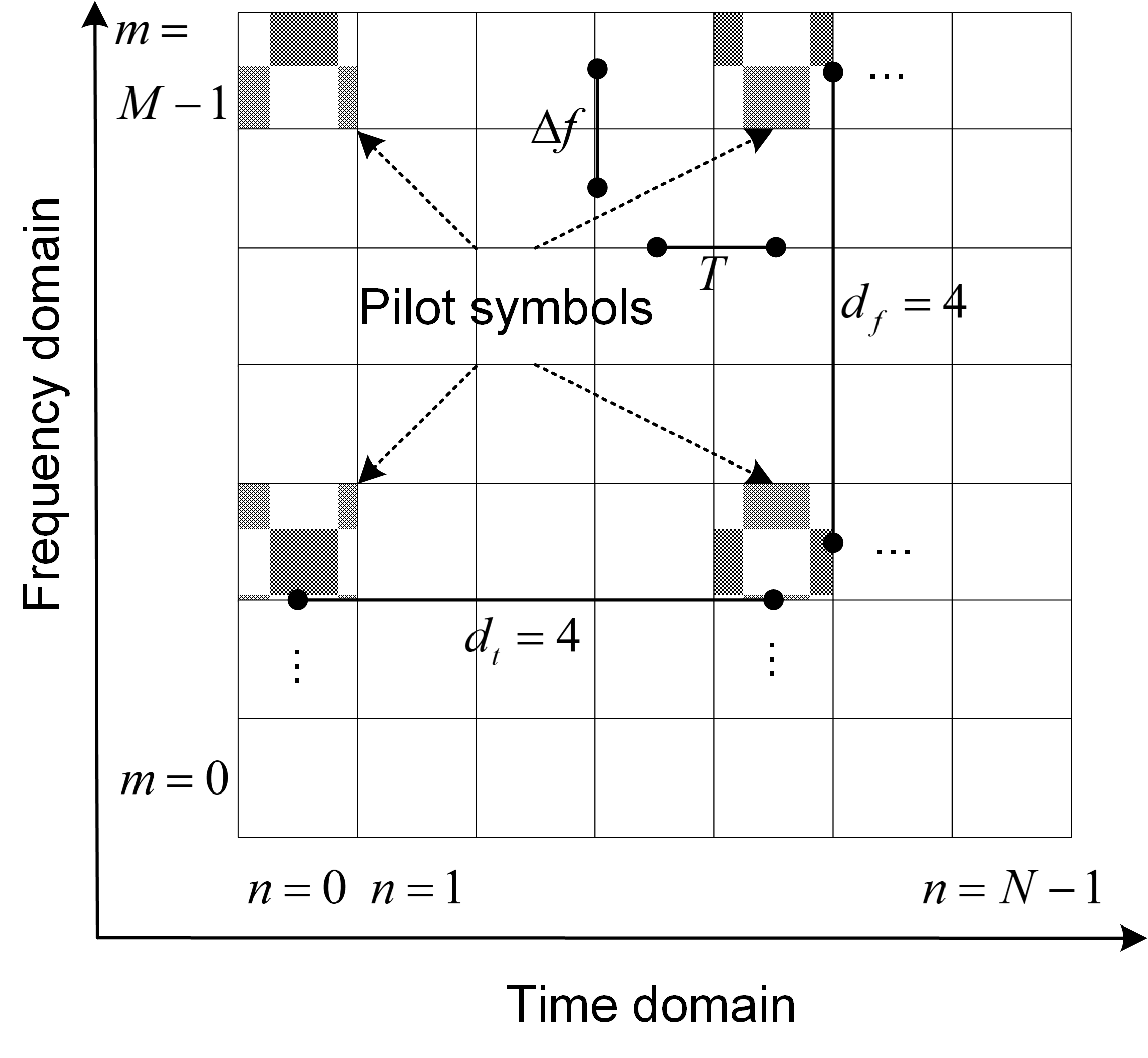}
  \vspace{-1em}
  \caption{Lattice-type pilot arrangement for the OFDM system.}\label{FIG1}
  \vspace{-1em}
\end{figure}
\par

\subsection{Existing Classical CTF Estimation Methods}
Two classical approaches are commonly used to estimate the CTF ${h}_{\rm TF}[m,n]$ in OFDM systems. These approaches are also applicable for the proposed pilot design. Specifically, both methods start by estimating ${h}_{\rm TF}[m,n]$ for the pilot symbols, i.e., $h_{\rm TF}^{\rm Pilot}[m,n]$ in (\ref{CTF_Dis}), based on the least square principle as:
\begin{equation}
{\hat h}^{\rm Pilot}_{\rm TF}[m,n] = {{\bf y}_{\rm TF}[m,n]}/{{\bf x}_{\rm TF}[m,n]}.
\label{LS}
\end{equation}
Then, they interpolate $\hat h_{\rm TF}^{\rm Pilot}[m,n]$ for the estimation of CTF at data symbols $h_{\rm TF}[m,n]$ with $ m \!\in\! {\bf S}^{1}_{M}\backslash{\bf S}^{d_f}_{M},{n\!\in\! {\bf S}_N^1 \backslash {\bf S}_N^{d_t}}$ \cite{b2}.\par

In the first approach, linear interpolation is typically conducted \cite{b2}. The time domain interpolated CTF becomes:
\begin{equation}
\begin{aligned}
{\hat h}_{\rm TF}[m,n] & \!=\!({\hat h}^{\rm Pilot}_{\rm TF}[m,(\tilde n+1)d_t] -{\hat h}^{\rm Pilot}_{\rm TF}[m,\tilde nd_t]) \frac{n'}{d_t}\\
& +{\hat h}^{\rm Pilot}_{\rm TF}[m,\tilde nd_t],
\end{aligned}
\label{LI}
\end{equation}
where $\tilde n = \lfloor \frac{n}{d_t} \rfloor$, $n' = {\rm mod}(n,{d_t})$, and similar interpolation can be performed 
along the frequency domain. We note that the CTF estimation approach in (\ref{LI}) is simple and effective for slow-fading channels, but its accuracy may be compromised in doubly selective channels.\par

The second approach estimates CTF of data symbols using the MMSE principle, given by \cite{b3}:
\begin{equation}
\begin{aligned}
\!{{\rm vec}(\hat{h}_{\rm TF})}\!=\!\mathbf{R}_{1}\!\left(\!\mathbf{R}_{2}\!+\!\sigma_w^2 {\mathbb E}\left(\mathbf{x}_{\rm TF} \mathbf{x}_{\rm TF}^{\mathrm{H}}\right)^{-1}\!\right)\!^{-1}\!{\rm vec}({\hat h}^{\rm Pilot}_{\rm TF}),
\end{aligned}
\label{MMSE}
\end{equation}
where ${\rm vec}(\bf{\Omega})$ denotes to vectorizing operation of the matrix $\bf{\Omega}$ and $\sigma_w^2$ is the variance of AWGN. Despite its estimation accuracy even under doubly selective channels \cite{b3}, this approach requires exact statistical information of the CTF, including $\mathbf{R}_{1} = {\mathbb E}({{\rm vec}(\hat{h}_{\rm TF})}{{\rm vec}(\hat{h}_{\rm Pilot})^H})$ and $\mathbf{R}_{2} = {\mathbb E}({{\rm vec}(\hat{h}^{\rm Pilot}_{\rm TF})}{{\rm vec}(\hat{h}_{\rm TF}^{\rm Pilot})^H})$ in (\ref{MMSE}), which are difficult to obtain. Given the challenges associated with these two approaches, we propose a novel CTF estimation approach next.\par


\section{CTF Estimation Inspired by the Stationary CSF}
It is well understood that the CSF can be quasi-time-invariant during the stationary region, which is significantly longer than the channel coherence region \cite{matz}. Leveraging this, we first estimate the CSF $h_{\rm DD}[k,l]$ for all $k\in{\bf S}^{1}_{N},l\in{\bf S}^{1}_{M}$ from the discrete CTF $h_{\rm TF}^{\rm Discrete}[m,n]$. We then derive the CTF for data symbols $h_{\rm TF}[m,n]$ with ${n\!\in\! {\bf S}_N^1 \backslash {\bf S}_N^{d_t}}, m \!\in\! {\bf S}^{1}_{M}\backslash{\bf S}^{d_f}_{M}$ based on the estimated $h_{\rm DD}[k,l]$.\par

Based only on ${h}^{\rm Discrete}_{\rm TF}[m,n]$ in (\ref{CTF_Dis}), an estimate for CSF can be made, which we denote as $h^{\rm Periodic}_{\rm DD}\left[k,l\right]$:
\begin{equation}
\begin{aligned}
&h^{\rm Periodic}_{\rm DD}\left[k,l\right]= DFT_{N}\{IDFT_{M}\{{{h}^{\rm Discrete}_{\rm TF}[m,n] }\}\}\\
& {=} \! \sum\limits_{i \! =\!  1}^P \!{h_i}\!\! \! \underbrace{\sum\limits_{m' \!=\! 0}^{\frac{M}{d_f} \!-\! 1} \! \frac{1}{\! \sqrt {\tilde M}}{{e^{ \!-j2\pi m'd_f\frac{{\left( {{l_i}\! -\! l} \right)}}{M}}}}}_{{\cal R}^{\rm Periodic}_{\rm delay}(l_i,l)}\underbrace{\!\sum\limits_{n' \!=\! 0}^{\frac{N}{{{d_t}}}\!-\!1}\! \frac{1}{\! \sqrt {\tilde N}} {{e^{ j2\pi n'd_{t}\frac{{\left( {{k_i} \!-\! k} \right)}}{N}}}} }_{{\cal R}^{\rm Periodic}_{\rm Doppler}(k_i,k)},
\end{aligned}
\label{CSF_dis}
\end{equation}
where $k\in{\bf S}^{1}_{N}$, $l\in{\bf S}^{1}_{M}$, and $\tilde M$ and $\tilde N$ are scaling factors that will be discussed later. We note that $h^{\rm Periodic}_{\rm DD}\left[k,l\right]$ in (\ref{CSF_dis}) is two-dimensional periodic with periods $\frac{M}{d_f}$ and $\frac{N}{d_t}$ over delay and Doppler domains, and thus:
\begin{equation}
h^{\rm Periodic}_{\rm DD}\left[{k+\tilde n{N}/{d_t}},{l+\tilde m{M}/{d_f}}\right] = h^{\rm Periodic}_{\rm DD}\left[k,l\right],
\label{CSF_per}
\end{equation}
for $k \in\{-\frac{N}{2 d_t},-\frac{N}{2 d_t}+1,...,\frac{N}{2 d_t}-1\}$, $l \in {\bf S}^1_{M/d_f}$, $\tilde{n} \in {\bf S}^1_{d_t}$, and $\tilde{m} \in {\bf S}^1_{d_f}$. Now we observe that for all $k \in {\bf S}^1_{N}$ and $l \in {\bf S}^1_{M}$, the $h_{\rm DD}^{\rm Periodic}[k,l]$ in (\ref{CSF_dis}) estimated from pilot symbols, is similar to $h_{\rm DD}[k,l]$ in (\ref{CSF}) estimated from $NM$ symbols. We will leverage this similarity in the following analysis.\par
We first note that the delay and Doppler components in (\ref{CSF_dis}), i.e., ${{\cal R}^{\rm Periodic}_{\rm delay}(l_i,l)}$ and ${{\cal R}^{\rm Periodic}_{\rm Doppler}(k_i,k)}$, respectively, have similar expressions, generally differing by a phase rotation of $\pi$. Thus, without loss of generality, we focus on ${\cal R}^{\rm Periodic}_{\rm Doppler}$ of (\ref{CSF_dis}) in the following discussion, which is expanded based on the geometric series expansion as follows:
\begin{equation}
\begin{aligned}
&{{\cal R}^{\rm Periodic}_{{\rm{Doppler}}}}(k_i,k)\!=\!\begin{cases} \frac{N}{d_t\sqrt{\tilde N}}, & k \!=\! k_i, \\
\frac{1}{\sqrt{\tilde N}}\frac{({{e^{j2\pi{{\left( {{k_i} - k} \right)}}}} - 1})}{({{e^{j2\pi {d_t}\frac{{\left( {{k_i} - k} \right)}}{N}}} - 1})}, & k \!\ne\! k_i.
\end{cases}
 \end{aligned}
\label{Period8}
\end{equation}
Similarly, the Doppler sampling component corresponding to the {\it original CSF} in (\ref{CSF}) can be written as:
\begin{equation}
\begin{aligned}
&{{\cal R}_{{\rm{Doppler}}}}(k_i,k)\!=\!\begin{cases}\sqrt{N}, & k \!=\! k_i, \\
\frac{1}{\sqrt{N}}\frac{({{e^{j2\pi{{\left( {{k_i} - k} \right)}}}} - 1})}{({{e^{j2\pi \frac{{\left( {{k_i} - k} \right)}}{N}}} - 1})}, & k \!\ne\! k_i.
\end{cases}
 \end{aligned}
\label{Origin9}
\end{equation}

Next, we will describe our channel estimation approach, beginning with the on-grid Doppler case, followed by the off-grid Doppler case.

\subsection{Estimation of CSF and CTF with On-grid Doppler}
For on-grid Doppler, i.e., $k_i$ is an integer, (\ref{Period8}) and (\ref{Origin9}) can be simplified to:
\begin{equation}
\small
\begin{aligned}
&{{\cal R}^{\!\rm Periodic}_{\!{\rm{Doppler}}}}\!(\!k_i,k\!)\!=\!\begin{cases} \!\frac{N}{d_t\sqrt{\!\tilde N}}\!, \!\!& k \!=\! k_i, \\
0,\!\! & k \!\ne\! k_i.
\end{cases}
 \end{aligned}\!\!\!
\begin{aligned}
&{{\cal R}_{\!{\rm{Doppler}}}}\!(\!k_i,k\!)\!=\!\begin{cases}\!\sqrt{N}\!, \!\!& k \!=\! k_i, \\
0, \!\!& k \!\ne\! k_i.
\end{cases}
 \end{aligned}
  \label{On-or}
\end{equation}
From (\ref{On-or}), to ensure that one period of ${{\cal R}^{\rm Periodic}_{{\rm{Doppler}}}}(k_i,k)$ exactly matches ${{\cal R}_{{\rm{Doppler}}}}(k_i,k)$ for $k \in\{-\frac{N}{2 d_t},...,\frac{N}{2 d_t}-1\}$, we can set $\tilde{N} = {N}/{d_t^2}$ and require $k_i \in [-{N}/(2d_t),{N}/(2d_t)-1]$. Similarly, as ${{\cal R}^{\rm Periodic}_{\rm delay}(l_i,l)}$ in (\ref{CSF_dis}) follows the similar principles as ${{\cal R}^{\rm Periodic}_{{\rm{Doppler}}}}(k_i,k)$, we need $\tilde{M} = {M}/{d_f^2}$ and $l_i \in {\bf S}^{1}_{{M}/d_f}$. Based on these, we arrive at the following theorem.
\begin{theorem}
For $k \in\{-\frac{N}{2 d_t},-\frac{N}{2 d_t}+1,...,\frac{N}{2 d_t}-1\}$ and $l \in {\bf S}^1_{M/d_f}$, $h^{\rm Periodic}_{\rm DD}\left[{k},{l}\right]$, could exactly represent the original CSF, if delay $\tau_i$ and Doppler $\nu_i$ are: i) on-grid with respect to the resolutions $\frac{1}{M\Delta f}$ and $\frac{1}{NT}$, respectively, and limited to ii) $\nu_{i} \in [-\frac{1}{2d_tT},\frac{1}{2d_tT}-\frac{1}{NT}]$ and $\tau_{i} \in [0,\frac{1}{d_f\Delta f}-1]$.
\end{theorem}
\begin{remark}
We emphasize that the periodic CSF $h_{\rm DD}^{\rm Periodic}[k,l]$ in (\ref{CSF_dis}) and Theorem 1 adhere to the Nyquist sampling rule. Given the continuous domain $h(\tau,\nu)$ is the Fourier pair of continuous domain $h(t,f)$, sampling $h(t,f)$\footnote{In OFDM systems, sampling the received time-windowed signals effectively results in $h(t,f)$ being sampled.} to obtain the discrete CTF $h_{\rm TF}[m,n]$ and $h_{\rm TF}^{\rm Discrete} [m,n]$ leads to their Fourier pair's $h_{\rm DD}[k,l]$ and $h_{\rm DD}^{\rm Periodic}[k,l]$ being periodic. Given $h_{\rm TF}[m,n]$ is obtained by sampling $h(t,f)$ at $t = nT$ and $f = m\Delta f$, $h_{\rm DD}[k,l]$ is periodic w.r.t. $1/T$ and $1/\Delta f$. Given $h_{\rm TF}^{\rm Discrete} [m,n]$ is obtained by sampling $h(t,f)$ at $t = nd_tT$ and $f = md_f\Delta f$, $h_{\rm DD}^{\rm Periodic}[k,l]$ is periodic w.r.t. $1/(d_tT)$ and $1/(d_f\Delta f)$. In the case of sampling $h(t,f)$ to obtain $h_{\rm DD}^{\rm Periodic}[k,l]$, to achieve sampling without aliasing, the sampling rate must exceed the support of $h(\tau,\nu)$.
\end{remark}
\begin{remark}
We conclude from Theorem 1 that there is no resolution reduction when estimating the original CSF using one period of $h^{\rm Periodic}_{\rm DD}\left[k,l\right]$ under the compactness conditions, i.e., $\nu_{i} \in [-\frac{1}{2d_tT},\frac{1}{2d_tT}-\frac{1}{NT}]$ and $\tau_{i} \in [0,\frac{1}{d_f\Delta f}-1]$. Therefore, the discrete CTF of pilot symbols in (\ref{CTF_Dis}) provides sufficient information for estimating the original CSF, while the CTF of $NM$ symbols is not necessarily required.
\end{remark}
Under the condition specified by Theorem 1, we can get the estimation of $h_{\rm DD}\left[k,l\right]$ for $k\in{\bf S}^1_{N}$ and $l \in {\bf S}^1_{M}$:
\begin{equation}
\begin{aligned}
&{\hat h}_{\rm DD}\left[{k},{l}\right] \!\!=\!\! \begin{cases} \hat h^{\rm Periodic}_{\rm DD}\left[k,l\right],\!\!\!\! & k \!\in\!\{\!-\frac{N}{2 d_t}\!,...,\!\frac{N}{2 d_t}\!-\!1\!\},\ \!l\! \in \!{\bf S}^1_{M/d_f},\\
0, \!\!\!\!& {\rm elsewise},
\end{cases}
\end{aligned}
\label{EQU10}
\end{equation}
where $\hat h^{\rm Periodic}_{\rm DD}\left[k,l\right]$ is the estimated $h^{\rm Periodic}_{\rm DD}\left[k,l\right]$ corresponding to (\ref{LS}). Finally, based on the estimated CSF ${\hat h}_{\rm DD}\left[{k},{l}\right]$ in (\ref{EQU10}), the estimation of $h_{\rm TF}[m,n]$ for data symbols can be straightly obtained as:
\begin{equation}
\hat h_{\rm TF}[m,n]= DFT_{N}\{IDFT_M\{\hat h_{\rm DD}[k,l]\}\}.
\label{CTF_FINALL}
\end{equation}
\subsection{Estimation of CSF and CTF with Off-grid Dopplers}
We note that due to limited $NT$, Dopplers of multipath are generally off-grid. In this case, the estimated ${\hat{\cal R}_{{\rm{Doppler}}}}(k_i,k)$ in (\ref{EQU10}) differs from (\ref{Origin9}). The difference is given by:\begin{equation}
\small
\begin{aligned}
&\Delta {{\cal R}_{{\rm{Doppler}}}} (k_i,k)= \\
& \begin{cases}
\begin{aligned}&\frac{1}{{\sqrt N }}{e^{j\left( {{k_i} - k} \right)\pi \frac{{(N - 1)}}{N}}}\frac{{\sin (\pi \left( {{k_i} - k} \right))}}{{\sin (\pi \frac{{\left( {{k_i} - k} \right)}}{N})}} \\
&- \frac{{{d_t}}}{{\sqrt N }}{e^{j\left( {{k_i} - k} \right)\pi {d_t}\frac{{(\frac{N}{{{d_t}}} - 1)}}{N}}}\frac{{\sin (\pi \left( {{k_i} - k} \right))}}{{\sin (\pi {d_t}\frac{{\left( {{k_i} - k} \right)}}{N})}}\end{aligned}, \!\!\!\!& k \!\in\!\{\!-\!\frac{N}{2 d_t}\!,\!...,\!\frac{N}{2 d_t}\!-\!1\!\},\\
\frac{1}{\sqrt N}\frac{{{e^{j2\pi \left( {{k_i} - k} \right) }}}- 1}{{{e^{\!j2\pi \frac{{\left( {{k_i} - k} \right)}}{N}}}}-1},& {\rm elsewise}. \end{cases}
\end{aligned}
\label{EQU12}
\end{equation}

\begin{remark}
  The error in (\ref{EQU12}) can also be understood through the aliasing effect of the Nyquist sampling criterion. Specifically,  when $k_i$ is off-grid, ${{\cal R}_{{\rm{Doppler}}}}(k_i,k)$ in (\ref{Origin9}) is non-zero for $k\in{\bf S}^1_{N}$. Given this, ${{\cal R}_{{\rm{Doppler}}}}(k_i,k)$ exceeds beyond the region defined by the sampling rates $1/(d_tT)$ and $1/(d_f\Delta f)$, that enables to obtain ${h}^{\!\rm Discrete}_{\rm TF}[m,n]$ from $h(t,f)$. This phenomenon results in DD domain aliasing in $h^{\rm Periodic}_{\rm DD}\left[k,l\right]$ and causes errors inevitably.
\end{remark}

To address the estimation error indicated by (\ref{EQU12}), we now demonstrate that it is still feasible to estimate the original CSF directly using one period of $h^{\rm Periodic}_{\rm DD}\left[k,l\right]$. Denote $k_i$ and $l_i$ as $k_i = k^{\rm I}_i + k^{\rm F}_i$ and $l_i = l^{\rm I}_i$, respectively, where $k^{\rm I}_i$ and $l^{\rm I}_i$ are integers and $k^{\rm F}_i \in (-0.5,0.5)$. For $k_i,k \in\{-\frac{N}{2 d_t},-\frac{N}{2 d_t}+1,...,\frac{N}{2 d_t}-1\}$, $l_i,l \in {\bf S}^1_{{M}/{d_f}}$, we can rewrite (\ref{CSF_dis}) as:
\begin{equation}
\small
\begin{aligned}
&h^{\rm Periodic}_{\rm DD}\left[k,l\right]= \sum\limits_{i = 1}^P {h_i}{{\cal R}^{\rm Periodic}_{\rm delay}(l_i,l)}\\
& \times \underbrace{\frac{{{d_t}}}{{\sqrt N }}{e^{j\left( {{k^{\rm I}_i + k^{\rm F}_i} - k} \right)\pi {d_t}\frac{{(\frac{N}{{{d_t}}} - 1)}}{N}}}\frac{{\sin (\pi \left( {{k^{\rm I}_i + k^{\rm F}_i} - k} \right))}}{{\sin (\pi {d_t}\frac{{\left( {{k^{\rm I}_i + k^{\rm F}_i} - k} \right)}}{N})}}}_{{\cal R}^{\rm Periodic}_{\rm Doppler}(k_i,k)}.
\end{aligned}
\label{CSF_dis_ir}
\end{equation}

Based on the property of the function ${{\rm sin}(x)}/{{\rm sin}(x/N)}$, we first note that $|{{\cal R}^{\rm Periodic}_{\rm Doppler}(k_i,k)}|$ in (\ref{CSF_dis_ir}) is maximum when $-0.5<k_i-k<0.5$, i.e., $k_i-0.5<k_i^{\rm I}+k_i^{\rm F}<k_i+0.5$. That is, $k = k_{0} = k^{\rm I}_i$ maximizes $|{{\cal R}^{\rm Periodic}_{\rm Doppler}(k_i,k)}|$. Next, we can find $k = k_0'$ which maximizes $|{{\cal R}^{\rm Periodic}_{\rm Doppler}(k_i,k)}|$ for $k\ne k^{\rm I}_i$, where $|k_0'-k_0| = 1$ holds, and obtain \cite{shi}:
\begin{equation}
\small
\begin{aligned}
\!&\frac{{\left| {{{\cal R}^{\rm Periodic}_{\rm Doppler}}({k_i},{k_0})} \right|}}{{\left| {{\cal R}^{\rm Periodic}_{\rm Doppler}({k_i},k_0')} \right|}}\!=\!\left| {\frac{{\sin (\pi \left( {k_i^{\rm{F}}} \right))}}{{\sin (\pi {d_t}\frac{{k_i^{\rm{F}}}}{N})}}} \right|\left| {\frac{{\sin (\pi {d_t}\frac{{\left( {{k_0} \!+\! k_i^{\rm{F}} \!-\! k_0'} \right)}}{N})}}{{\sin (\pi \left( {{k_0} \!+\! k_i^{\rm{F}} \!-\! k_0'} \right))}}} \right|\\
& = \left| {\frac{{\sin (\pi {d_t}\frac{{\left( {k_i^{\rm{F}} - k_0' + {k_0}} \right)}}{N})}}{{\sin (\pi {d_t}\frac{{{k^{\rm F}_{{i}}}}}{N})}}} \right|\overset{(a)}{\approx} \left| {\frac{{\left( { {k^{\rm F}_{{_i}}}- {k'_0} + {k_0}} \right)}}{{{k^{\rm F}_{{_i}}}}}} \right|,
\end{aligned}
\label{EQU15}
\end{equation}
where equation $\overset{(a)}{\approx}$ is obtained based on the assumption that $N$ is relatively large. Therefore, we can estimate ${k^{\rm F}_{{_i}}}$ as:
\begin{equation}
\small
\hat k_i^{\rm{F}} = \frac{{\left| {{\cal R}^{\rm Periodic}_{\rm Doppler}({k_i},{k_{0}'})} \right|\left( {{k_0'} - {k_{0}}} \right)}}{{\left| {{\cal R}^{\rm Periodic}_{\rm Doppler}({k_i},{k_0})} \right| + \left| {{\cal R}^{\rm Periodic}_{\rm Doppler}({k_i},{k_{0}'})} \right|}},
\label{EQU16}
\end{equation}
and the estimation of $k_i$ as $\hat k_i = k_0 + \hat k_i^{\rm{F}}$. Similarly, when $|{{\cal R}^{\rm Periodic}_{\rm delay}(l_i,l)}|$ is maximum for $l = l_0$, we can estimate $l_i$ as $\hat l_i = l_0$. Under the assumption that there is one path at the maximum at each resolvable delay grid, Algorithm \ref{Alg1} is proposed to estimate the original CSF \cite{shi}. 
\setlength{\textfloatsep}{10pt}
\begin{figure}[t]
\vspace{-1em}
\begin{algorithm}[H]
\caption{Original CSF Estimation Algorithm} 
\label{Alg1}
\begin{algorithmic}[1]
\REQUIRE ~~\\ 
Estimated $\hat h^{\rm Periodic}_{\rm DD}\left[k,l\right]$; Number of Multipath $\hat P$;\\
\ENSURE ~~\\ 
$\hat h_i$, $\hat l_i$, $\hat k_i$ for $i\in\{1,...,\hat P\}$;
\FOR{$i \in [1,\hat P]$}
\STATE $(k_0,l_0) \!=\! \mathop {{\mathop{\rm argmax}\nolimits} }\limits_{k \in\!\{\!-\!\frac{N}{2 d_t},\!-\frac{N}{2 d_t}\!+\!1,...,\frac{N}{2 d_t}\!-\!1\},
l \in {\bf S}^1_{M/d_t}} \!\left|\!{\hat h_{{\rm{DD}}}^{{\rm{Periodic}}}\left[ {k},{l} \right]} \!\right|$;\
\STATE $k'_0=\underset{k \in \{k_0-1,k_0+1\}}{\operatorname{argmax}}\left| {\hat h_{{\rm{DD}}}^{{\rm{Periodic}}}\left[ {{k},{l_0}} \right]} \right|$;\
\STATE $\hat k_i = k_0 + \frac{{\left| {\hat h_{{\rm{DD}}}^{{\rm{Periodic}}}\left[ {k'_0},{l_0} \right]}  \right|}\left( {{k_0'} - {k_{0}}} \right)}{{\left| {\hat h_{{\rm{DD}}}^{{\rm{Periodic}}}\left[ {{k_0},{l_0}} \right]} \right| + \left| {\hat h_{{\rm{DD}}}^{{\rm{Periodic}}}\left[ {k'_0},{l_0} \right]} \right|}}$; $\hat l_i = l_0$;\
\STATE $\hat h_i = \frac{{\hat h_{{\rm{DD}}}^{{\rm{Periodic}}}\left[ {{k_0},{l_0}} \right]}}{{{\cal R}^{\rm Periodic}_{\rm delay}(\hat l_i,l_0)}{{\cal R}^{\rm Periodic}_{\rm Doppler}(\hat k_i,k_0)}}$;\
\STATE ${\hat h_{{\rm{DD}}}^{{\rm{Periodic}}}\left[ {{k},{l}} \right]} = 0$ for $l = l_0$ and $k \in {\bf S}^1_{M}$;
\ENDFOR
\end{algorithmic}
\end{algorithm}
\vspace{-1em}
\end{figure}

In Step 1, in addition to $\hat h^{\rm Periodic}_{\rm DD}\left[k,l\right]$ in (\ref{EQU10}), we will input the estimation of the number of multipath $\hat P$, which can be observed through an energy threshold \cite{shi}. From Steps 2 to 4, $\hat k_i$ is estimated based on (\ref{EQU15}) and (\ref{EQU16}). $\hat l_i$ is estimated in Step 4 based on the property of ${{\cal R}^{\rm Periodic}_{\rm delay}(l_i,l)}$. Next, in Step 5, we obtain $\hat h_i$ based on $\hat k_i$, $\hat l_i$, and (\ref{CSF_dis_ir}). In Step 6, the CSF corresponding to the $i$-th path is nulled to estimate parameters of the $(i+1)$-th path. Finally, $\{\hat h_i, \hat k_i, \hat l_i\}$ for $\hat P$ multipath is obtained. Based on Algorithm 1, the estimation of the origin CSF with off-grid Doppler is obtained as:
\begin{equation}
\begin{aligned}
&{\hat h}_{\rm DD}[{k},{l}]{=}\!\!\sum\limits_{i {=} 1}^{\hat P} {{{\hat h}_i}\!\sum\limits_{m {=} 0}^{M {-} 1} \!\frac{1}{\sqrt M}{{e^{ - j2\pi m\frac{{\left( {{\hat l_i} - l} \right)}}{M}}}\sum\limits_{n {=} 0}^{N {-} 1} \!\frac{1}{\sqrt N}{{e^{j2\pi n\frac{{\left( {{\hat k_i} {-} k} \right)}}{N}}}} } }.
\end{aligned}
\label{CSF_EQU17}
\end{equation}
Finally, the CTF for pilot symbols can be obtained based on (\ref{CTF_FINALL}). Meanwhile, we clarify that Algorithm 1 is also designed based on the compactness of the original CSF to detect the maximum values of $|{{\cal R}^{\rm Periodic}_{\rm Doppler}(k_i,k)}|$ and $|{{\cal R}^{\rm Periodic}_{\rm Delay}(l_i,l)}|$ in (\ref{CSF_dis_ir}). Therefore, we present the following theorem.
\begin{theorem} Based on the observation of one period of $h^{\rm Periodic}_{\rm DD}\left[{k},{l}\right]$, the original CSF can be exactly estimated if $\nu_{i} \in [-\frac{1}{2d_tT},\frac{1}{2d_tT}-\frac{1}{NT}]$ and $\tau_{i} \in [0,\frac{1}{d_f\Delta f}-1]$. \end{theorem}
\begin{remark}
According to Theorems 1 and 2, we note that the proposed discrete CTF based CSF estimation model is feasible for both on-grid and off-grid channel conditions.
\end{remark}
\section{Numerical Results}
In this section, we evaluate the performance of the proposed CSF-inspired CTF estimation method. We consider the extended vehicular A channel environment \cite{raviteja}, with a maximum speed of mobile terminals of $v = 250$ \text{km/h}, a carrier frequency of $f_{\rm c} = 2.1$ \text{GHz}, and $\Delta f = 1/T = 15$ \text{kHz}. We set $M = 128$ and $N = 64$ \footnote{Accordingly, we can obtain $NT\nu_{\max} = 2.06 \ll 64$ to make (\ref{OFDMIOSIMPLE}) valid.}, with $d_t = 4$ and $d_f = 4$ to ensure $\nu_{i} \in [-\frac{1}{2d_tT},\frac{1}{2d_tT}-\frac{1}{NT}]$ and $\tau_{i} \in [0,\frac{1}{d_f\Delta f}-1]$. Besides, 4-QAM modulation is considered, and the energies of pilot and data symbols in TF domain are the same.

\begin{figure}
\centering
\subfigure[]{
\includegraphics[width=.24\textwidth,height = 125pt]{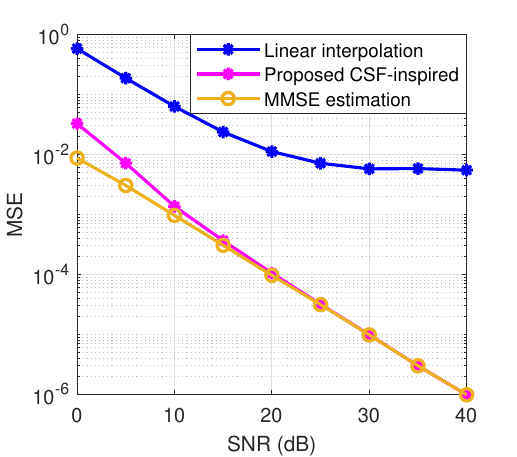} 
}\subfigure[]{
\includegraphics[width=.24\textwidth,height = 125pt]{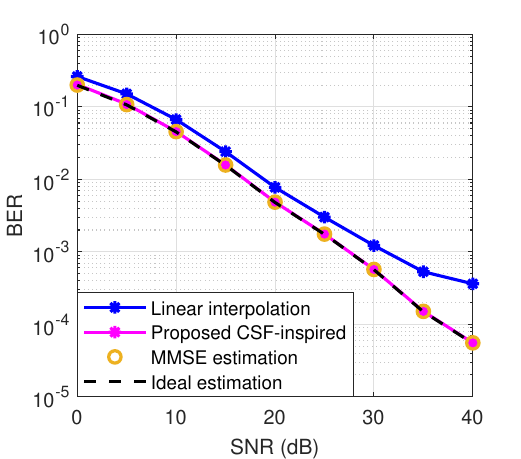} 
}
\DeclareGraphicsExtensions.
\vspace{-1em}
\caption{Performance of CTF estimation with on-grid delays and on-grid Dopplers: (a) MSE and (b) BER.}\label{FIG2}
\vspace{-1.5em}
\end{figure}
\begin{figure}
\centering
\subfigure[]{
\includegraphics[width=.24\textwidth,height = 125pt]{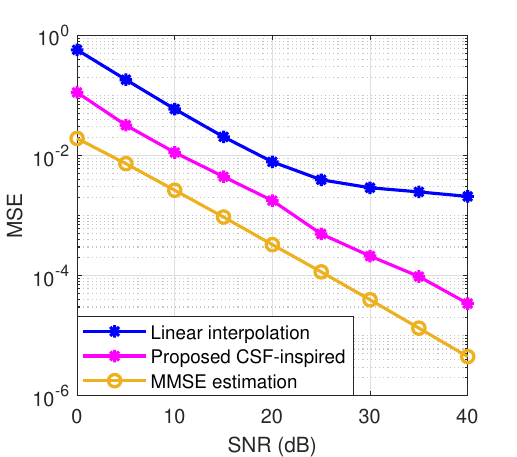} 
}\subfigure[]{
\includegraphics[width=.24\textwidth,height = 125pt]{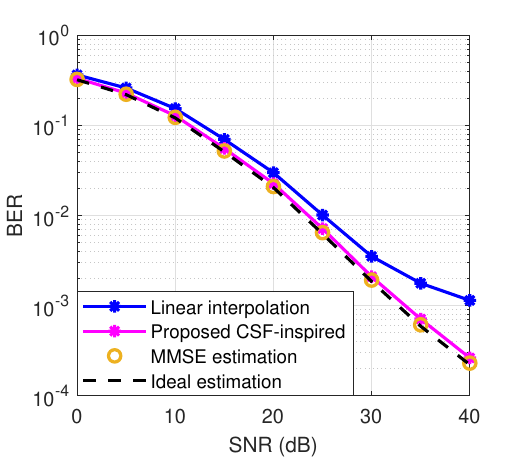} 
}
\DeclareGraphicsExtensions.
\vspace{-1em}
\caption{Performance of CTF estimation with on-grid delays and off-grid Dopplers: (a) MSE and (b) BER.}\label{FIG3}
\vspace{-1em}
\end{figure}

In Figs. \ref{FIG2} and \ref{FIG3}, the mean square error (MSE) of CTF estimation and bit error rate (BER) with a single-tap equalizer for different signal-to-noise ratios (SNRs) under both on-grid and off-grid Doppler scenarios are plotted. First, the performance of the proposed CSF-inspired CTF estimation approach is presented. For comparison, we include: i) the joint least square estimation and linear interpolation approach, given in (\ref{LS}) and (\ref{LI}), and ii) the MMSE channel estimation approach, as indicated in (\ref{MMSE}). Additionally, the ``ideal estimation'' case, where the CTF is perfectly known, is included as a benchmark.\par

As shown in Figs. \ref{FIG2}(b) and \ref{FIG3}(b), we first observe that the CSF-inspired CTF estimation achieves BER performance close to that of the ideal channel estimation case. This confirms the validity of our proposed Theorems 1 and 2, as well as the effectiveness of the estimation methods.\par

Second, the proposed design outperforms the linear interpolation approach in terms of both MSE and BER. Notably, the linear interpolation method reaches error floors for both MSE and BER in the high SNR region, as it is not well-suited for fast-fading CTF. In contrast, our proposed design effectively leverages the time-invariant property of the CSF, achieving accurate CTF estimation and enabling reliable equalization.\par

Third, as shown in Fig. \ref{FIG2}(a), the MSE of the proposed design is nearly identical to that of the linear optimal MMSE estimator in the high SNR region. Additionally, in both Fig. \ref{FIG2}(a) and Fig. \ref{FIG3}(a), we observe that the MSE of the proposed design is higher in the off-grid Doppler scenario compared to the on-grid scenario, due to the approximations used by Algorithm \ref{Alg1}. Nevertheless, the BER performance of the proposed design remains nearly identical to that of the MMSE estimator in both scenarios. Notably, while the MMSE estimator depends on channel statistical information in (\ref{MMSE}), the proposed design does not. The complexity of the proposed design is ${\cal O}(N \log N + M \log M + P)$, as determined by (\ref{CSF_per}), (\ref{CSF_EQU17}), and Algorithm 1, compared to the MMSE estimator's complexity of ${\cal O}(\frac{N^3}{d_t^3} \frac{M^3}{d_f^3})$ in (\ref{MMSE}). These observations indicate that by leveraging the stationary properties of the doubly selective channel, the proposed CSF-inspired CTF estimation approach offers near-optimal performance with significantly lower complexity and without requiring prior information.
\section{Conclusions and Future Works}
This paper addressed CSF-inspired CTF estimation for OFDM systems in high-mobility scenarios. First, we presented theorems and methods for CSF estimation based on the discrete CTF of pilot symbols predefined in the TF domain. The CTF was then derived from the estimated CSF. Simulation results validated the proposed approach, demonstrating MSE and BER performance comparable to the optimal MMSE estimator, with the advantage of not requiring channel statistical information and lower complexity. This study highlights how leveraging the stationary CSF can improve performance of the commercial OFDM in high-mobility environments. Future work will focus on enhancing the proposed design by accounting for ICI and exploring other pilot patterns.

\end{document}